\documentclass{ws-procs9x6}
\begin{document}

\title{CONFINEMENT OF COLOR: RECENT PROGRESS.
\footnote{\uppercase{T}his work is partially supported by MIUR Progetto Teoria delle
interazioni fondamentali.}}

\author{Adriano Di Giacomo}
\address{Dipartimento di Fisica, Universit\`a di Pisa,I.N.F.N. Sezione di Pisa,\\
Via Buonarroti 2, 56127 Pisa, Italy\\
E-mail: adriano.digiacomo@df.unipi.it}  

\maketitle

\abstracts{
Recent progress done in Pisa on the subject is presented. It is shown
that dual superconductivity of the vacuum or absence of it is an
intrinsic property of QCD vacuum, independent of the choice of the
abelian projection. The order of the deconfining phase transition in $N_f=2$
QCD is studied as a key to understand the mechanism of confinement.}

\section{Dual superconductivity: dependence on the abelian projection.}
The mechanism of confinement by dual superconductivity of the vacuum
\cite{tHM}requires the identification in QCD of a U(1) fiber
bundle, which has to be color gauge invariant and color singlet if color
is not broken by monopole condensation in the vacuum. The procedure for
that is known as "Abelian Projection" \cite{tH81}.

A disorder parameter is then introduced\cite{ZAK,refr},
$\langle\mu\rangle$ which is the vev of a magnetically charged operator
$\mu$ .
$\langle\mu\rangle
\neq 0$ means Higgs breaking of the magnetic U(1) gauge symmetry, or dual
superconducticity, and is expected to hold in the confined phase $T<T_c$.
For $T  > T_c$ (deconfined phase) instead the $U(1)$ is not broken and
$\langle\mu\rangle=0$.

  It can be instructive to start from the SU(N) Higgs model, which is
defined by the Lagrangean
\begin{equation}
L = -\frac{1}{4}Tr\{G_{\mu\nu} G_{\mu\nu}\} - Tr\{D_\mu\phi^\dagger D_\mu \phi\}
- V(\phi)
\end{equation}
  The Higgs field $\phi$ is a N by N matrix and transforms in the adjoint
representation. In the Higgs phase  $\langle\phi\rangle$ = $\phi^a\neq0$.
The index $a$ labels different minima of $V(\phi)$.

The relevant quantity for monopole solutions is\cite{tH74}
\begin{equation}
F_{\mu\nu} =
Tr\left\{ \phi G_{\mu\nu}\right\} - \frac{i}{g}
Tr\left\{ \phi \left[D_\mu\phi, D_\nu\phi\right]\right\}
\end{equation}
A theorem can be proved on $F_{\mu\nu}$ \cite{SUN}:

A necessary and sufficient condition (i) for the cancellation 
of bilinear terms in $A_\mu A_\nu$
between the
two terms on the rhs of eq(2) and (ii) for the validity of the Bianchi
identities $\partial_\mu F_{\mu\nu}^*=0$ is that    
\begin{equation}
\phi^a(x) = U(x) \phi^a_{diag} U^\dagger(x) 
\end{equation}
with
\begin{equation}
\phi_{diag}^a = 
 diag\left(
\overbrace{\frac{N-a}{N},..,\frac{N-a}{N}}^{a},
\overbrace{-\frac{a}{N},..,-\frac{a}{N}}^{N-a}\right)
\end{equation}
and $U(x)$ an arbitrary gauge transformation.
  The little group of $\phi^a_{diag}$ is $SU(a) \otimes SU(N-a)\otimes U(1)$. It identifies
a symmetric space\cite{W}. If $L_0$ is the corresponding Lie algebra ,
$L$ is the Lie algebra of $SU(N)$ ,and $L_1 = L - L_0$ ,
$[L_0,L_0]\subset L_0$,
$[L_0,L_1]\subset L_1$, $[L_1,L_1]\subset L_0$. It can be shown that all
possible symmetric subspaces of $SU(N)$ have the form eq(3).

Viceversa, if the Higgs field $\phi$ belongs to the adjoint representation
then the breaking identifies a symmetric space,i.e. $\phi^a$ has the form
eq(3)\cite{MCH}.

For a Higgs field of the form eq(3) one has identically
\begin{equation}
F^a_{\mu\nu} =
\partial_\mu Tr\left\{ \phi^a A_\nu\right\} - 
\partial_\nu Tr\left\{ \phi^a A_\mu\right\}
-\frac{i}{g}
Tr\left\{ \phi^a \left[\partial_\mu\phi^a, \partial_\nu\phi^a\right]\right\}
\end{equation}
 $F^a_{\mu\nu}$  is gauge invariant by construction. In the unitary gauge     
$\partial_\mu\phi^a=0$, the
second term of eq(5) vanishes and
\begin{equation}
F^a_{\mu\nu} = \partial_\mu Tr(\Phi^a A_\nu) - \partial_\nu Tr(\Phi^a A_\mu)
\end{equation}
assumes an abelian form. The transformation to the unitary gauge is
called  abelian projection.

  Expanding the diagonal part of the fields 
$A^\mu_{diag}  =  \alpha^i A^i_\mu$, in terms of the roots
\begin{equation}
\alpha^i = diag(0,0,0\ldots\stackrel{i}{1},\stackrel{i+1}{-1},0\ldots 0)
\qquad tr\{\alpha^i \phi^a_{diag}\} = \delta^{ia}
\end{equation}
gives
\begin{equation}
F^a_{\mu\nu} = \partial_\mu   A^a_\nu - \partial_\nu  A^a_\mu
\end{equation}
A monopole solution exists in the $SU(2)$ subspace spanned by the diagonal
elements $i$ and $i+1$, of the form of the $SU(2)$
solution\cite{tH74,P75}.
For this solution  $E_i=F_{0i}=0$ and 
\[ \vec H = \frac{2\pi}{g}\frac{\vec r}{r^3} +\hbox{Dirac string}\]
 whence the name of monopole to the soliton.

A magnetic current can be defined
\begin{equation}
J_\nu^a =\partial_\mu F^{a *}_{\mu\nu}
\end{equation}
Bianchi identities require that
$ J^a_\mu=0$  but it can be different from zero in a
compact formulation like lattice, in which the Dirac string is invisible.
In any case
\begin{equation}
 \partial_\mu J^a_\mu= 0
\end{equation}
  $N-1$ $U(1)$ magnetic symmetries are thus defined, which are topological
symmetries, which do not correspond directly to invariances of the
lagrangean. In QCD they will be the magnetic symmetries which will
eventually be Higgs broken in the confined phase, producing dual
superconductivity.

The construction of the magnetically charged operators $\mu^a$ which
will provide the disorder parameters for dual superconductivity is
the following\cite{refr}. The denomination "disorder" comes from
statistical mechanics and means that $\langle\mu^a \rangle$ is expected
to be different from zero in the confined phase,which is strong coupling
(disordered), and to be equal to zero in the ordered phase.
\begin{equation}
\mu^a(\vec x,t) =
e^{
i\int d^3\vec y\,Tr\left(\phi^a_{}\vec E(\vec y,t)
\right)
\vec b_\perp(\vec x -\vec y)
}
\end{equation}
with
\begin{equation}
\vec\nabla\vec b_\perp = 0\quad, \vec\nabla\wedge \vec b_\perp
= \frac{2\pi}{g}\frac{\vec r}{r^3} +\hbox{Dirac string}
\end{equation}
$\mu^a$ is gauge invariant if $\phi^a$ belongs to the adjoint
representation.

In the abelian projected gauge   $\phi^a = \phi^a_{diag}$
$Tr(\vec E(y,t)\phi^a_{diag}) = \vec E^a$
and
\begin{equation}
\mu^a(\vec x,t) =
\exp\left\{i\int d^3\vec y\,{\vec E}_\perp^a(\vec y,t)
\vec b_\perp(\vec x -\vec y)\right\}
\end{equation}
where only the transverse part of $\vec E^a$ survives in the convolution with
$\vec b_\perp$.

In whatever quantization scheme $\vec E^a_\perp$ is the conjugate momentum to
$\vec A^a_\perp$, and hence
\begin{equation}
\mu^a(\vec x,t) | {\vec A}^a_\perp(\vec y,t)\rangle
=  | {\vec A}^a_\perp(\vec y,t) + \vec b_\perp(\vec x - \vec y)\rangle
\end{equation}
$\mu^a$ creates a monopole in the U(1) generated by $\alpha^a$ in the
abelian projected gauge.

The construction can be repeated unchanged in the Coulomb phase, in spite
of the fact that there are no monopoles as solitons, by taking any
$\phi^a(x)  = U(x)\phi^a_{diag}U(x)\dagger$ in the adjoint representation.
If $U(x)$ is defined as the gauge transformation which diagonalizes the
Higgs field  $\phi(x)$, then
$\phi^a(x)$ is diagonal with $\phi(x)$ , but any other choice provides an
abelian projection : for example $\phi^a(x)$ can be diagonal in the
maximal abelian gauge.

$\mu^a$ depends on the choice of U(x)
\begin{equation}
\mu^a(\vec x,t) =
e^{
i\int d^3\vec y\,Tr\left(\phi^a_{diag}U^\dagger(\vec y,t)\vec E(\vec y,t)
U(\vec y,t)
\right)
\vec b_\perp(\vec x -\vec y)
}
\end{equation}

If $U(x)$ does not depend on the gauge field configuration, when computing
correlators of $\mu^a$ it can be reabsorbed by a change of variables
which leaves the measure invariant\cite{D},and
\begin{equation}
\mu^a(\vec x,t) =
e^{
i\int d^3\vec y\,Tr\left(\phi^a_{diag}\vec E(\vec y,t)
\right)
\vec b_\perp(\vec x -\vec y)
}
\end{equation}
  All memory of $U(x)$ has disappeared, and $\langle\mu^a\rangle\neq0$ or
$\langle\mu^a\rangle=0$ are statements independent on the abelian
projection.

If $U(x)$ depends on $A_\mu(x)$ in general the measure is not invariant and
a non trivial jacobian can appear after gauge transformation by $U(x)$, so
that the correlation functions of $\mu^a$ , and in particular its vev are
projection dependent.
  However if the number density of monopoles is finite the operator
$\mu^a(x)$ defined by eq(16) will create a monopole in all abelian
projections, since the gauge transformation to any abelian projection
will be continuous in a neighbourhood of x and will preserve topology.
  Hence, if the number density of monopoles is finite the statement
$\langle\mu^a\rangle\neq0$ and $\langle\mu^a\rangle=0$ are abelian projection
independent\cite{D,DP}.

Dual superconductivity (or non-superconductivity) of the vacuum is
an intrinsic property, independent on the particular choice of the
abelian projection.

In QCD there are no Higgs fields, but, as discussed above, any operator
in the adjoint representation $O(x)$ can provide an abelian projection,
  in the sense that the operator $U(x)$ of eq(2) can be chosen as the
one which diagonalizes $O(x)$. Again, if the number density of monopoles
is finite dual supercondutivity is an intrinsic property, independent of
the choice of the abelian projection.
  The density of monopoles can be estimated by looking at the distribution
of the difference of the eigenvalues of any operator in the adjoint
representation on the sites of a lattice. The location of monopoles
coincides indeed with such zeros\cite{tH81}. We have studied that
distribution on samples of lattice configurations , with different
lattice spacings and for a number of operators. A typical distribution is
shown in fig~1, which refers to the Polyakov line as operator, $10^3$
configurations on a $16^4$ lattice ,quenched SU(3) and $\beta= 6.4$
The number of sites on which there is a monopole is zero.

\begin{figure}
\includegraphics[angle=270,scale=0.4]{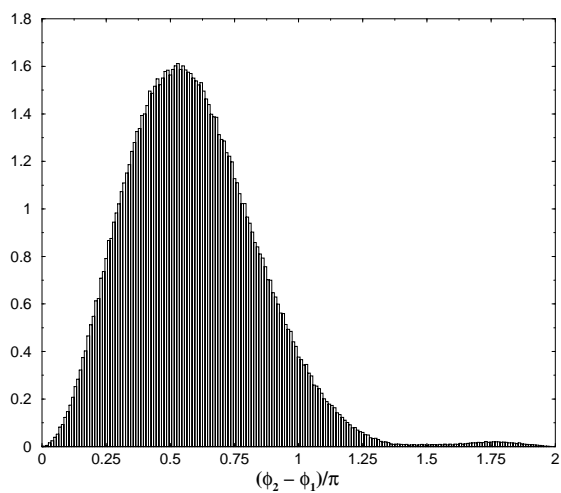}
\caption{An example of probability distribution of the difference of the two highest
eigenvalues of the phase $\Phi$ of the  Polyakov line $e^{i\Phi}$, at the lattice
sites. $SU(3)$ gauge group, $\beta = 6.4$, lattice $16^4$, $10^3$ configurations.}
\end{figure}

\begin{figure}
\includegraphics[scale=0.4]{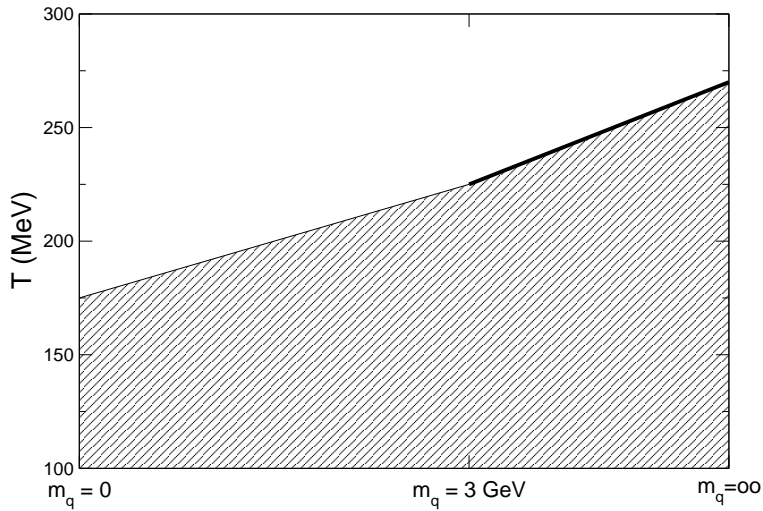}
\caption{The phase diagram of two flavor QCD.}
\end{figure}

An independent test can be made by numerical comparison of the order
parameter for different abelian projections, which confirms the
independence of dual superconductivity on the choice of the abelian
projection.\cite{ref22}

  The measurement of the disorder parameters $\langle\mu^a\rangle$ in the
quenched case works as follows\cite{refr}. Instead of $\langle\mu^a\rangle$
one determines the quantity $\rho^a\equiv \frac{d}{d\beta} ln \langle\mu^a\rangle$.
$\rho^a$ is a susceptibility. At the deconfining transition where
$\langle\mu^a\rangle$ has a sharp drop $\rho^a$ has a negative peak. A
phase transition can only take place in the infinite volume
limit\cite{LY}. As the volume increases the drop of $\langle\mu^a\rangle$
becomes sharper and sharper and the negative peak of $\rho$ higher and
higher. Since \cite{refr}
\begin{equation}
\langle\mu^a\rangle = exp(  \int_0^\tau \rho^a(\tau') d\tau')
\end{equation}
and for $T<T_c$ $\rho^a$ becomes volume independent within numerical
errors with increasing volume, one concludes that for $T<T_c$
$\langle\mu^a\rangle\neq0$, implying dual superconductivity.
For $T>T_c$  $\rho \sim -|c| N_s$  with $N_s$ the spatial extension of the
lattice, implying that $\langle\mu^a\rangle=0$ strictly in the
thermodynamical limit (normal vacuum).
  In the critical region the correlation length $\xi$ goes large ,the
ratio of the lattice spacing a to $\xi$, $a/\xi$ can be put to zero,
and then the disorder parameter only depends on the ratio $N_s/\xi$
\begin{equation}
\langle\mu^a\rangle=\tau^\delta\phi(N_s/\xi)
\end{equation}
whence the scaling law follows
\begin{equation}
\rho^a/N_s^{1/\nu}= f(\tau N_s^{1/\nu})
\end{equation}
  In particular the peak height scales as $N_s^{1/\nu})$ whence $\nu$
can be determined. The result is consistent\cite{refr} with the values
obtained by use of the Polyakov line\cite{K,SU3}, or
$\nu=.62$ for pure gauge $SU(2)$ (3d ising universality class) , and .33
for $SU(3)$, first order transition.

\section{Two flavor QCD.}
  In quenched theory one uses the Polyakov criterion to define confinement,
which refers to the static potential between a quark and an antiquark.

The
order parameter is $\langle L\rangle$ the Polyakov line : when
$\langle L\rangle=0$ the potential grows linearly with distance, when
$\langle L\rangle\neq0$ it goes to a constant. 

Of course one should in
principle show that confinement defined in this way implies the absence
of any colored particle in asymptotic states, which is not easy to do,
but the criterion is reasonable anyway. As shown above it fits with
identifying confinement with dual superconductivity of the vacuum.

In the presence of dynamical quarks $Z_3$ symmetry is explicitely broken
and
$\langle L\rangle$ cannot be an order parameter. Moreover string breaking
is expected to occur: the potential energy stops growing with distance,
due to the instability for production of quark antiquark pairs,even if
there is confinement. At $m_q=0$ there is chiral symmetry, which is known
to be spontaneously broken at zero temperature, the pseudoscalar mesons
being the Goldstone particles. The symmetry is restored at some
temperature $T_c$, where the order parameter $\langle\bar\psi\psi\rangle$ goes
to zero. It is not known what is the relation between chiral symmetry
breaking and confinement. In any case at $m_q\neq0$ chiral symmetry is
explicitely broken and  $\langle\bar\psi\psi\rangle$ is not an order
parameter either.
   The situation for $N_f=2$ $m_u=m_d=m$ is depicted in fig~2.

A number of susceptibilities can be measures on the lattice as functions
of the temperature T at given $m$ ( The susceptibility of
$\langle\bar\psi\psi\rangle$, that of $\langle L\rangle$,the specific
heat\cite{KL,FU}). All of them show a peak at the same $T(m)$,
which defines the curve in the phase diagram of fig~2. By convention the
region below that curve is called confined, the region above it
deconfined.

A renormalization group analysis can be made\cite{PW} of the chiral
transition assuming that the Goldstone particles are the relevant degrees
of freedom at the transition , with the following result. For $N_f=3$ the
chiral transition is first order and such is the transition at $m\neq0$.
For $N_f=2$ if the anomaly of the U(1) axial current vanishes below $T_c$
the transition is first order and such is the transition at $m\neq0$; if
instead the anomaly persists up to $T_c$ the transition is second order
in the universality class of O(4) and the line at $m\neq0$ is a crossover.
Lattice data are not yet conclusive on this issue, but for some reason
the second possibility is usually assumed to be true.

  A possible criterion for confinement could be dual superconductivity of
the vacuum, which is already valid in the quenched case. Indeed the disorder
parameter $\langle\mu^a\rangle$ can equally well be defined in the
presence of dynamical quarks. Lattice simulations show\cite{ref3} in fact
that
$\langle\mu^a\rangle$ is non zero below the critical line of fig~2, and
is stricly zero above it in the thermodynamical limit. Of course in
principle one should show that dual superconductivity implies absence of
colored particles in asymptotic states, which is not trivial to do: but
the situation is not different from that of the Polyakov criterion, as
discussed in the previous section.

  A finite size scaling analysis around $T_c$ can be performed to get
information on the order of the phase transition. The issue is very
relevant to understand confinement. Indeed if the determination gives
a result consistent with what is obtained by studying the specific heat
a legitimation results for $\langle\mu^a\rangle$ as an order parameter
and for dual superconductivity as a mechanism of confinement.
  Preliminary data\cite{PICA} indicate that the chiral transition is
first order and certainly not in the universality class of O(4). A careful
analysis is being completed, which will  give an unambiguous answer to the
question.  A careful analysis of the anomaly around $T_c$ is also on the
way to check consistency with ref.\cite{PW}.
  Some details on the analysis. A new scale is present in the problem with
respect to the quenched case. Eq(18) now reads
\begin{equation}
\langle\mu^a\rangle = \phi(N_s/\xi, mN_s^{y_h})
\end{equation}
  The problem can be reduced to a single scale by choosing masses and
sizes such that $mN_s^{y_h}=constant$ , assuming for $y_h$ alternatively
the value corresponding to O(4) universality class ($y_h=2.49$) or the
value for a first order transition $y_h=3$. For the same values different
susceptibilities and the specific heat can be determined, and the
critical indices can be measured consistently. This program is being completed.

Thanks are due   J.M. Carmona, L. Del Debbio, M.
D'Elia, B. Lucini, G. Paffuti  and C. Pica who collaborated to obtain  the
results presented here .


\begin{thebibliography}{0}
\bibitem{tHM} G.'tHooft in High Energy Physics  EPS Conference Palermo
1975, A.Zichichi ed.; S.Mandelstam, {\it Phys.Rep.} {\bf 23C} , 245 (1976).
\bibitem{tH81}G.'tHooft , {\it Nucl.Phys.}  {\bf B190} ,455 (1981).
\bibitem{ZAK} A. Di Giacomo, {\it Acta Physica Polonica}  {\bf B25}, 227 (1994).
\bibitem{refr} L. Del
Debbio, A. Di Giacomo, G. Paffuti, {\it Phys. Lett.}  {\bf B349} , 513 (1995); A. Di
Giacomo, B. Lucini, L. Montesi, G. Paffuti, {\it Phys. Rev.} {\bf D61} , 034503
(2000); A. Di Giacomo, B. Lucini, L. Montesi, G. Paffuti, {\it Phys. Rev.}{\bf D61},
 034504 (2000).
\bibitem{tH74} G.'tHooft, {\it Nucl.Phys.}  {\bf B79}, 276 (1974).
\bibitem{SUN}L. Del Debbio, A. Di Giacomo, B. Lucini, G.
Paffuti hep-lat/0203023.
\bibitem{W} S.Weinberg {\it The quantum theory of fields}, Chapt.19,  Cambridge
Univ.Press (1990).
\bibitem{MCH} L. Michel {\it Rev.Mod.Phys.}  {\bf 52}, 617 (1980).
\bibitem{P75}A. M. Polyakov, {\it JETP Lett.} {\bf 20}, 894 (1974).
\bibitem{D}A. Di Giacomo, hep-lat/0206018
\bibitem{DP} A. Di Giacomo, G. Paffuti, {\it The abelian projection revisited}
(LATTICE 2003) hep-lat/0309019
\bibitem{LY}C.N.Yang, T.D. Lee, {\it  Phys.Rev.} {\bf 87}, 404 (1952).
\bibitem{ref22}J.M. Carmona, M. D'Elia, A. Di Giacomo, B.
Lucini, G. Paffuti, {\it  Phys.Rev.} {\bf D64} , 114507 (2002).
\bibitem{K} J. Fingberg, U. M. Heller, F. Karsch, {\it Nucl.
Phys.} {\bf B392}, 493 (1993).
\bibitem{SU3}M. Fukugita, M. Okawa, A. Ukawa, {\it Phys.Rev. Lett.} {\bf 63}, 1768
(1989).
\bibitem{PW}R. Pisarski and R. Wilczek, {\it Phys. Rev.} {\bf D29},
338 (1984).
\bibitem{KL}F. Karsch and E. Laermann, {\it Phys. Rev.} {\bf D50},
6954 (1994)
\bibitem{FU} S. Aoki et al. (JLQCD collaboration), {\it Phys.
Rev.} {\bf  D57}, 3910 (1998).
\bibitem{ref3} J.M. Carmona, M. D'Elia, L. Del Debbio, A.
Di Giacomo, B. Lucini, G. Paffuti, {\it Phys. Rev.}
{\bf D66}, 011503 (2002).
\bibitem{PICA}J.M. Carmona, M. D'Elia, L. Del Debbio, A. Di Giacomo, B.
Lucini, G. Paffuti, C. Pica ,{\it Deconfining transition in two-flavor QCD.}
(LATTICE2003) hep-lat/0309035.
\end{thebibliography}
\end{document}